# A NOVEL FRAMEWORK FOR ELECTRONIC GLOBAL HEALTH RECORD ACCESS

Nael A.H AbuOun[1], Ayman Abdel-Hamid[1], and Mohamad Abou El-Nasr[2]

[1]College of Computing and Information Technology
[2]College of Engineering and Technology
Arab Academy for Science, Technology, and Maritime Transport, Alexandria, Egypt

## ABSTRACT

*When most patients visit physicians in a clinic or a hospital, they are asked about their medical history and related medical tests' results which might not exist or might simply have been lost over time. In emergency situations, many patients suffer or sadly die because of lack of pertinent medical information. Patient's Health information (PHI) saved by Electronic Medical Record (EMR) could be accessible only by a hospital using their EMR system. Furthermore, Personal Health Record (PHR) information cannot be solely relied on since it is controlled solely by patients. This paper introduces a novel framework for accessing, sharing, and controlling the medical records for patients and their physicians globally, while patients' PHI are securely stored and their privacy is taken into consideration. Based on the framework, a proof of concept prototype is implemented. Preliminary performance evaluation results indicate the validity and viability of the proposed framework.*

## 1. INTRODUCTION

Most people have their health information scattered in many different places. Health Information could be kept at the patient' household (Papers, or electronic format), at their physician or therapist office, anywhere they've been hospitalized before, or at any personal website. Hence, many problems occur when collecting all health information of the patient in a reasonable time. The best solution to solve such problems is to keep all health information together in one place along with all the medical actors (physicians, hospitals, patients, Ministries of Health (M.O.H), World Health Organization (W.H.O)) and specifying privileges for accessing what is needed when it is needed through the right actor.

Medical Records of a patient consist of medical and personal health information, such as medical history, care or treatments received, any test results done before, diagnoses from physicians, and any medications taken. In many places such record is still paper-based and is susceptible for loss or damage.

The use of the PHI as a digital format went through three successive techniques starting with the Electronic Medical Record (EMR), then the Electronic Health Record (EHR), and finally the Personal Health record (PHR). The first technique, Electronic Medical Record, is defined as "*a digital version of a paper chart that contains all of a patient's medical history from one practice.*

*An EMR is mostly used by providers for diagnosis and treatment*" [1]. The problem with EMR is that it is hospital dependent and thus is very much similar to the paper-based technique of saving medical data.

The second technique, Electronic Health Record, is defined as *"a longitudinal electronic record of patient health information generated by one or more encounters in any care delivery setting. Included in this information are patient demographics, progress notes, problems, medications, vital signs, past medical history, immunizations, laboratory data and radiology reports"* [2]. Estonia was the first country to implement the first nationwide Electronic Health Record (EHR) in the world [3], which was launched on December 17, 2008 [4].

EHR has many advantages over EMR for the medical care specialist which can be different from one system to another [5]. According to Centers for Disease (CDC) and Prevention's National Center for Health Statistics survey of 2011 in USA, 85% of physicians are satisfied with their EHRs system, and 74% reported that EHRs enhanced their overall patient care [6]. The largest integrated healthcare EHR in USA is VistA_EHR providing care to over 8 million veterans, employing 180,000 medical personnel and operating 163 hospitals, and over 800 clinics [7]. VistA_EHR is also implemented in Jordan (Oct. 2009) since it was proven as national-scale enterprise system capable of scaling to hundreds of hospitals [8]. UK spends over $24 Billion to have all patients with a centralized electronic health record by 2010, but the program was dismantled after the high cost [9]. Canada started the EHR system in 2004 under the name MyHealthAlberta for Alberta province allowing health professionals to view patient provincial medication profiles and a selection of local laboratory test results [10].

Since both EMR and EHR were run only by hospitals or medical specialists, patients have no access to their PHI outside hospitals in the majority of EHRs. Consequently, a new technique was found called Personal Health Record (PHR) defined as "*a set of computer-based tools that allow people to access and coordinate their lifelong health information and make appropriate parts of it available to those who need it*" [11] [12].

PHR is portable and is kept with the patient and contains medical lifelong information. It should not be restricted by file formats or other local issues. PHR gave the patient the ability to share their PHI with other medical care centres and the ability to control their record. However, patients' PHI was still hard to be accessed in emergency situations. If the patient went in a comma, he/she won't be able to inform physicians where his/her PHR is located, and thus it is useless in such cases. According to a conducted survey [13] in an ongoing research effort, more than 90% of physicians need more than five minutes to retrieve patient's record, and 71% of targeted physician says they did face situations like that and explained how hard it was to deal with such situations like that [13]. Other examples of PHR are explained in related work.

Consequently, the new idea of a Global Health Record was introduced. It started in 2004 with The USA Armed Forces Health Longitudinal Technology Application *"(AHLTA) is the military's electronic health record, marks a significant new era in healthcare for the Military Health System (MHS) and the nation. It is used by medical providers of the U.S. Department of Defense (DOD) since its initial implementation in January 2004. AHLTA is a services-wide medical and dental information management system*" [14].

This paper introduces a framework that merges the benefits of EMR and PHR into one medical record called Global Health Record (GHR). GHR is a new methodology for making the medical record globally accessible anytime anywhere, allowing physicians to access patients' information from one location quickly granting privacy to the patient's information with high level of security. In this framework, patients do have full control over their record and at the same time physicians are allowed to access that record and have the ability to modify it. All actors can access the framework from the same domain which redirects each user according to their geographical location to the nearest front end (Figure (1)). Users then get authenticated, after that the medical data is required from the backend server which is located in a different cloud controlled by the patients' ministry of health in his/her country. The communication between the different backbend's is controlled by the W.H.O. cloud which identifies the backend servers and publishes their public keys.

Patients, physicians, hospitals, and all medical actors can access the framework from any Internet browser. One advantage of the proposed framework is that it solves the problem of accessing and sharing the medical record of the patient in a global perspective having the correct information written by physicians, and also sharing all lab tests. According to a survey conducted to outline motivation and requirements of the proposed framework [13], more than 71% of physicians face problems in sharing patient data and face delay in receiving the data in the required time, also more than 85% of the physicians say that the data in the received records is not accurate [13]. Another advantage of the proposed framework is that it lets the patients add to their records and also control their information privacy as in PHR. Moreover, this framework helps physicians around the world in their researches or conducted studies on variant diseases and their transformations. This is performed without violating patients' privacy because only the medical information is provided to physicians with no ties to the patient personal information. According to the conducted survey more than 78% of physicians reported that it is hard to locate the case study in search studies [13].

To establish a successful and trusted global medical framework, a trusted global organization such as (W.H.O.) must manage it. More than 64% of physicians accepted the W.H.O. to manage a global medical system [13]. W.H.O. has many connections around the world that will make the implementation of such framework easier. W.H.O. can administrate the framework around the world and register all ministries of health around the world in the framework authorizing the communication between different countries. Figure (1) represents the abstract idea of the framework. It illustrates that each country has its own "GHR cloud" saving and managing its actors (Hospitals, Physicians, and Patients) who can access the system through the front end cloud for each country. Proper redirection to the local front end cloud (country-wise) can be performed through Domain Name System (DNS) redirection. The main management for the communication between the countries will be administered by the W.H.O. The proposed framework is designed to be implemented via cloud computing which decreases high expenses of hardware, using any type of Software as a Service (SaaS), or Platform as a Service (PaaS). More framework details will be presented in sections III and IV.

The rest of the paper is organized as follows. Section II discusses relevant related work. Section III introduces the framework's analysis, design, and comparison versus related work. Section IV presents proof of concept prototype and performance evaluation results. Finally, Section V concludes this paper and discusses future work.

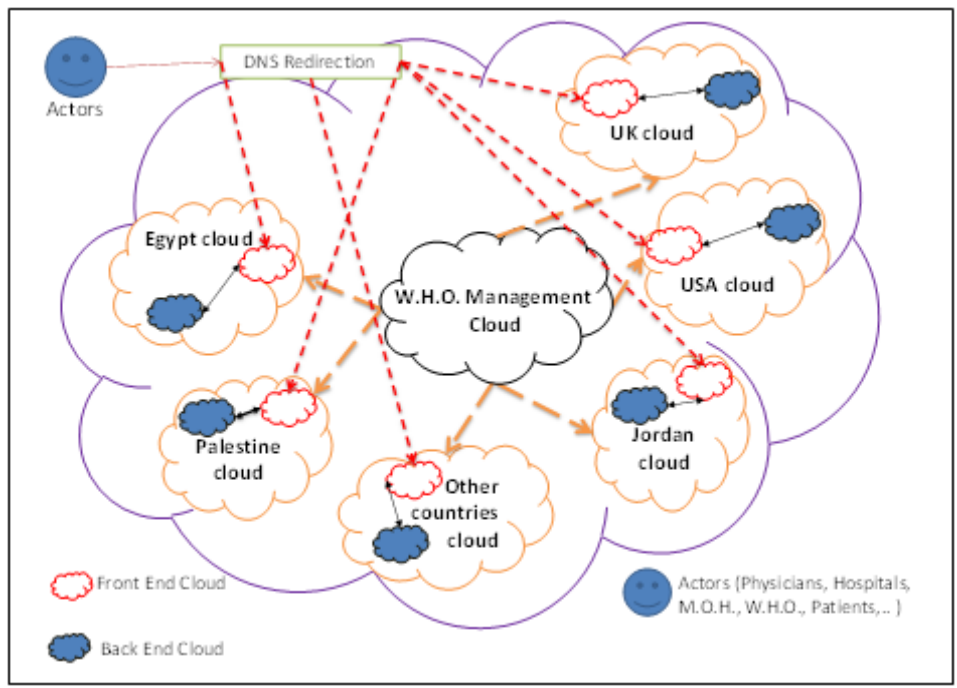

Figure 1 Abstract view of the framework

## 2. RELATED WORK

Many researches and projects were designed and implemented offering new ideas regarding EHR and PHR. However, the majority of such work was not designed for global sharing of the patients' health information. Moreover, security and privacy were not the most important aspects which caused data breaches, e.g., in 2011, 380 major data breaches were reported in United States, involving 500 or more patients' records listed on the website kept by the United States Department of Health and Human Services (HHS) Office for Civil Rights [15]. Moreover, 2013 was a busy year in medical data breaches according to HIPAA [16], when the five-hospital Riverside Health System in southeast Virginia announced that close to 1,000 of its patients were being notified of a privacy breach that continued for four years.

Many research work discussed health records and cloud computing. Scalable and Secure Sharing of Personal Health Records in Cloud Computing [17] proposed patient-centric framework of mechanisms for data access control to PHRs stored in semi-trusted servers over cloud. CAM: Cloud-Assisted Privacy Preserving Mobile Health Monitoring [18] designed a cloud assisted privacy preserving mobile health monitoring system to protect the privacy of the involved medical parties and their data. Sensor-Cloud: A Hybrid Framework for Remote Patient Monitoring [19] used Sensors to enable a patient monitoring system using cloud computing to monitor human health and share the information among doctors, care-takers, researchers, and pharmacies. SPOC: A Secure and Privacy-preserving Opportunistic Computing Framework for Mobile-Healthcare Emergency [20] introduced an efficient user-centric privacy access control, which is based on an attribute-based access control and a new privacy-preserving scalar product computation technique. In addition, it allows a medical user to decide who can participate in the opportunistic computing to assist in processing his overwhelming PHI data. A mobile Phone based homecare management system on the cloud [21] combined Hospital Information System (HIS) and mobile communications to establish a telemedicine homecare management system in long-term and sustainable health monitoring through the transmission of Multimedia Messaging Service (MMS).

A medical record is presented in many ways such as internet-based record, which has limitations relating to privacy and security concerns, information could be shared with others, may be hacked, and concerns about the internal design of the system and the way it works. Some examples of such systems include: MyPHR Internet-based PHR brought by AHIMA Foundation improves health [22], Microsoft Health Vault system [23], Google health which introduced in 2008 and cancelled in 2012, according to Brian Dolan there was 10 reasons why Google health was terminated [24] [25].

Some other research work explored the idea of using portable medical record such as mobile app, USB Flash, or on credit-card wallet CDs. Card that could be carried around on key chains, bracelets, watches, etc. Mycare card MyC2 [26] [27] in UK was developed with GUI software to control the insertion and modification of data. Privacy Preserving Portable Health Record (P^3HR) [28] proposes an architecture for USB flash device which provides data encryption, and strong multifactor authentication using biometrics, public key infrastructure to verify the credentials of the applicants. However, the device can be damaged and becomes useless in emergency situations when a patient's life depends on the data installed on it.

Portable Device PHR [29] proposes the structure for a portable PHR (pPHR) protected by password and dividing data to five different data types: Confidential, Normal, Transfer, Prescription, and Emergency and giving privileges according to user type to each section of data. However, concerns are in place if a patient deals with unregistered user in an emergency situation or the device gets lost or damaged. MyPHRMachines is PHR with full control of patient, which allows patients to build PHRs which are robust across the space and time dimensions, and share these data with different care institutions [30] [31] but still it is PHR which means patients adding all the information.

Other research and projects can be found on mobile apps for telemedicine such as Dhatri [32], a mobile Phone based homecare management system on the cloud [21], MTBC PHR, and SecEra PHR. Figure (2) shows the traditional personal health record system general design [33]. However, if any patient has accounts on all electronic records, this will not do him any good during an emergency where he can not provide information to physicians (a patient being in a comma for example). A physician does not know where to look for a PHR or they are not identified as trusted users.

In this paper, the proposed framework is going to address problems of accessing the medical record globally by making all patients and medical users securely communicate through one abstract system for sharing medical information while maintaining patients' privacy. Hence, the proposed framework presents a novel framework for electronic global health record access, allowing all medical actors to access the medical information from anywhere at any time in a secure manners (see sections III and IV).

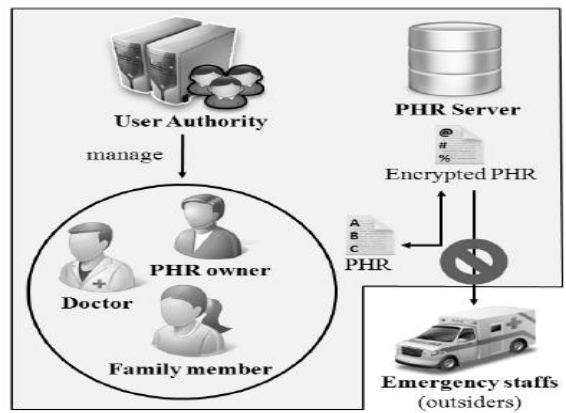

Figure (2) Traditional personal health record system [33]

## 3. ELECTRONIC GLOBAL HEALTH RECORD ACCESS FRAMEWORK

This section illustrates the framework's analysis and design, explaining each part of the proposed system, highlighting the privacy part for the medical information for patients, and the security for actor's access to the framework, and comparing between the proposed framework and other related work.

### 3.1 Framework Analysis

The framework's idea was formulated after searching for many EHRs, and PHRs presented in medical life today. To the best of our knowledge, moving to a global record was considered a problem since the privacy of data was hard to be saved and making sure that data was not going to be misused or exposed. This framework is designed to work at global level, considering the privacy and security of each record and trying to make it easy for medical users around the world by making one place as a portal gate for all medical actors. The globally trusted organization (W.H.O) will register all ministries of health around the world giving them proper authentication to access the framework, and enable them to create the sub-users in each country such as hospitals, physicians, and all the medical entities in that country as explained in figure (3). M.O.H. will be the controller of their sub-users by making sure of their registration, log-in information, and giving them the authentication to use the framework globally.

### 3.1.1   Framework Actors

Creating users will be done in hierarchical way. W.H.O creates accounts for each ministry of health, and then each ministry creates their sub-users, and so on as illustrated in figure (3). The following paragraphs will describe each type of actors discussing the privileges given to each one starting with the patients:

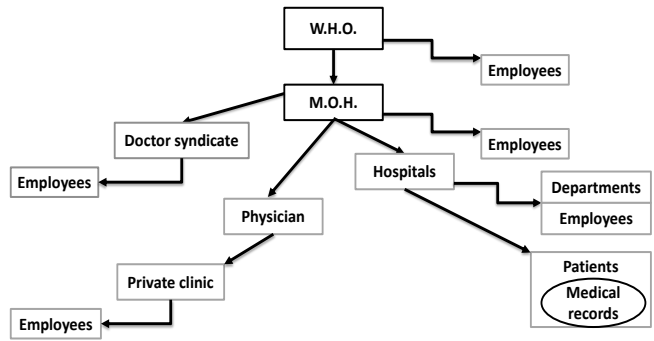

Figure (3) the Hierarchical registration process of framework entities

## Patients view

Each citizen in his/her country should have an ID Social Security Number in Palestine, Canada and USA (SSN) is 9 digit number, in Egypt 14 digit number, KSA and Sri Lanka 10 digit, and UAE 15 digit [34]. Since the SSN is differentiate between different countries, and could identify the patients; this framework started new ID for each patient to avoid any duplication or mistakes in patient ID, each patient will have global ID divided as shown in table (1).

Table (1) Patient ID

| Country ID | Province ID | City ID | Sequence ID |
|---|---|---|---|
| EG | ALX-02 | MAN-004 | FFFFFF |

The ID consists of 19 characters divided into four sections. The first section is the country ID which is predefined globally, then the province ID defined within the country and then city ID. Finally, the sequence ID is an automated sequence number. By using this record ID, the patient will be able to access their record through any browser connected to the Internet, even from outside their country, by using the username and password created at the time of the record creation.

Creating the records for the new born babies will be easy since the record will be created in hospital after birth, and then it should be verified by M.O.H by assigning a SSN to that record. Regarding current citizens, each one when he/she visits the hospital for the first time they will be assigned an ID which also will be verified by ministry of health later, and adding all the old data that was saved in the patient record by hospital employees. For making the record easy usable for patients there is the "Session Token – OTP (One Time Password)" which can be identified as a password for other actors access to patient record. No one knows it or can have it except the patient and it can be created only by the patient when he/she logs in to their account. The session Token is used when a patient visits a physician at a clinic to give them authority of seeing the full record with history and any other needed information.

## Physicians view

The physicians are going to modify patients' records and identify their medical situations to be accessible for any actor requiring this information. Since many countries rely on M.O.H. to identify the physicians working licenses (according to [13] more than 85% confirmed that M.O.H issues the licenses to physicians), this framework adopts a similar approach. M.O.H. will be able to create the physicians account giving them their ID as shown in Table (2). A physician ID consists of 19 characters. The area section is used for patient searches at any emergency situation to get closest physician to his/her location to ask for special visit. Hence, storing an area ID will enhance patient's search results.

Table (2) physician ID

| Country ID | Province ID | City ID | Area ID | Sequence ID |
|---|---|---|---|---|
| EG | ALX-002 | 003 | 003 | FFFFF |

**Hospitals view**

Each hospital can have their account by registering through the M.O.H, and each hospital can have their unique ID, as shown in Table (3).

When a patient visits a hospital to get medical treatment, the case added in his/her record will be tagged with physician, and hospital IDs. This can help when sometimes a physician could ask for more medical information about a specific patient from another physician by just looking to who diagnosed that case for the patient.

For each organization, we suggest to use a second way for authentication with the username and password to make it more secure in using the framework such as static IP with Virtual Private Network (VPN) connection, or hardware authentication. The second authentication for that organization will be identified by M.O.H when creating that organization's profile to make sure the usage of the username is located in the right organization, in order to organize the access to patients' data to save patient privacy.

Table (3) Hospital ID

| Org. identify Code | Country ID | Province ID | City ID | Sequence ID |
|---|---|---|---|---|
| HOS | EG | Alx-002 | 002 | 00001-01 |

**The Records**

The record is divided to a number of sections. Each section will have its own storage location in the cloud to allow for fault-tolerance. The main sections of a record will be medical information (cases, visits, and notes), insurance information, contact information, and any other information related to patient's record.

Each section contains specific data about a patient, for example the contact information section holds full information about a certain name and how to contact or reach the patient. Such information is secure and there is no way to link it to other record's parts except through patient's authentication. Other parts may be individually accessed by some other actors according to actor type and the privileges they had which is also controlled by patients.

Within the cloud, no linkability is available between records' sections. The only authorized actor to see the full information from all sections is the patient. Other actors do have the ability to communicate with the predefined section to that type of actor, for example a physician has the authority to only view the medical information according to where he/she is located. The patient has the privilege to give specific actors the access to different record sections.

By dividing the record, the patient will get the highest level of privacy, since the medical information is saved using a virtual ID that is not linkable to other parts of the record such as a patient's name. This implies that no one can access the patient's information, except who is having the privileges to access it. Such privileges can be changed by the patient through their account settings.

The framework will accept actor login from anywhere and will be able to differentiate between login locations by using a second authentication method for all registered organizations. Figure (4) illustrates the communication between physicians and cloud servers while trying to log in from a hospital network. Each type of actor will get privileges according to the location they are logging on from.

For protecting a patient's privacy, the framework limits other actors from accessing the record from anywhere, allowing a physician to make changes only when a patient is available in front of them in hospital or clinic by using biometric data (according to the survey in [13], more than 92% agreed to use the biometric data to identify the patients) or the Session Token of that patient. If a patient is not physically located with physician then it will allow access as shown in figure (5), for example when a patient visits a physician in a clinic. The key to view the record will be patient's fingerprint, or the Session Token. The record could be accessible for a specific physician when a patient adds that physician into a trusted list which could contain physicians, and hospitals trusted by that patient.

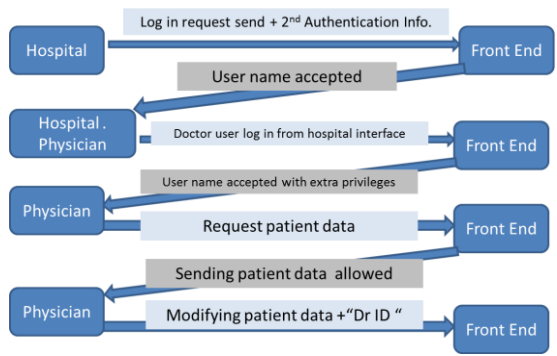

Figure (4) physician log in from hospital scenario

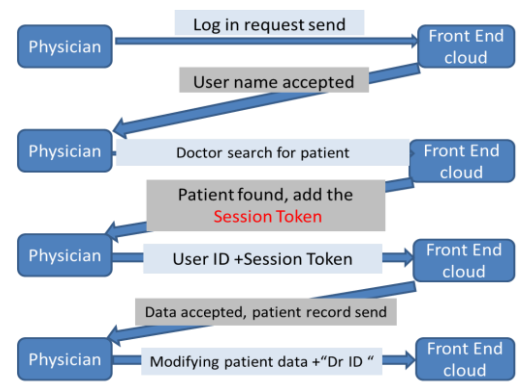

Figure (5) physician log in from clinic scenario

Each type of actor will be granted privileges according to the location they are logging on from. Physicians can access the framework from different locations such as hospital, clinic, home, mobile, etc… Each location of access gives the physicians different views of records as outlined in Table (4).

In some medical research studies, physicians would like to target specific diseases and send new medication or treatments about that case to the patients. In such case, a physician will not need the full information about that patient till the patient grants the privileges to that physician. They can just view diseases' situation and related medication and which physician did treat it. In order to get more information about that case from the patient, the physician can communicate with patient or physician who handles that case without knowing any other information about the patient till they are granted privileges by the patient. Patient data will be protected from non-authorized actor access and patient privacy will be maintained.

Table (4) physician's privileges according to location

| Physician | Hospital-Clinic | Patient | Privileges |
|---|---|---|---|
| **Present** | hospital | Not present | Searching by names for registered patient by that hospital only, and get full record. For patients not registered, just medical information and searching with IDs, or diseases |
| **Present** | hospital | Present | By using biometric data, or Session Token get full record. |
| **Present** | clinic | Not present | No ability to search by names except for patients who previously got treatment from same physician or patient records them as trusted physician. Ability to search by cases or diseases and just medical info can be viewed. |
| **Present** | clinic | Present | Biometric data, or session Token gives full record to physician |
| **Present** | Anywhere | Present | |
| **Present** | Anywhere | Not present | Searching only for treated patients by that physician, and others only by cases or diseases and only IDs available no names |

### 3.1.2    Security and Privacy

Many researches can be cited regarding the security and privacy of PHR [35], [36], [37], [38], [39], [40], and [41]. However, there are still several uncertain issues such as how to manage PHR information in emergency situations.

When talking about PHI or health records, security and privacy should be considered, the data privacy according to HIPAA deals with controlling who is authorized to access information. Hence, in order to maintain PHI privacy, security measures need to be implemented [42].

Medical Privacy can be identified as "*the practice of keeping information about a patient confidential. This involves both conversational discretion on the part of health care providers, and the security of medical records.*" [43]. Furthermore, according to HIPAA privacy of medical information "*could be the rule focuses on limiting the use and disclosure of sensitive (PHI)*" [44]. HIPAA (Section 165.501) provides for the protection of individually identifiable health information that is transmitted or maintained in any form or medium.

Technically, privacy can be classified as: anonymity, unlinkability, undetectability, unobservability, pseudonymity, or identity management [45], this framework focuses on anonymity and unlinkability of the patient's record.

Anonymity of a subject means that the subject is not identifiable within a set of subjects, the anonymity set. Unlinkability of two or more items of interest (IOI) (e.g., subjects, messages, actions ...) from attacker's perspective means that within the system (comprising these and possibly other items), attacker cannot sufficiently distinguish whether these IOIs are related or not [45].

Healthcare cloud applications share common security measures known as the CIA triad (confidentiality, integrity, and availability of data), and ownership of information, authentication, non-repudiation, patient consent and authorization. The aforementioned measures are outlined and handled in the proposed framework as follows.

a) **Confidentiality:** is defined in ISO-17799 as "*ensuring that information is accessible only to those authorized to have access*". Confidentiality can be achieved by access control and encryption techniques. The GHR record in this framework is saved in the cloud infrastructure using a digital ID that is just known to the internal service of cloud that is collecting the data from different locations to send it to an authenticated actor. Authenticated actors are predefined and each actor type will have specific privileges that are predefined as well, and controlled by the internal service. The data located in the cloud is encrypted. Communication between the entities will be also encrypted through asymmetric cryptography.

b) **Integrity**: Integrity means preserving the accuracy and consistency of data, and protecting information from being modified by unauthorized parties or modified in transit. According to HIPAA, the covered entity is required to provide corroboration of the integrity of the data through mechanisms such as checksums, CRCs (Cyclic Redundancy Checks), double keying, message authentication codes or digital signatures [46].

c) **Availability**: what is the benefit of confidentiality and integrity if the authorized users of information cannot access and use that data? The framework depends on cloud computing which means 24/7 availability and fault-tolerance allowing for zero data loss depending on cloud provisioning of virtual resources.

d) **Ownership of information**: Any created data is tagged with the creator username. Patient data is owned and managed by the patient himself.

e) **Authentication:** each user acquires a username and password for system access. A tracking service is available for each user to specify the locations they are able to access the system from. New locations should be confirmed by the user.

f) **Non-repudiation:** using the audit measure, the framework can record user activities and save it in the user's log. It also implies that one party of a transaction cannot deny having received a transaction nor can the other party deny having initiated a transaction.

g) **Patient consent and authorization**: each patient knows which physician(s) they do trust. From the system management board, they can control denying and/or allow sharing their information with others. In addition, they can monitor who are able to access their record and what record parts are accessible by that user.

## 3.2  Comparison versus Related Work

Related research work differs from the proposed framework in many aspects, which is explained as follows. In summary, Table (5) depicts the comparison between the proposed framework and related work.

a) Websites EHR, PHR: Susceptible for being hacked and unauthorized use. Data saved online susceptible for being used without user knowledge. Availability of data is controlled by the hosting organization which is not trusted. In some cases, when data is required in any emergency it maybe inaccessible. In addition, there is the problem of sharing data with other systems or care delivery organization when required.

b) Portable PHR: Could have good availability of data but problem to deal with in some cases when data is encrypted, and unable to decrypt it. Data saved in the portable device may have some inaccurate information since it is created by the patient. Susceptible for being damaged or lost. In addition, there is the problem of sharing data with other systems or care delivery organization when required.

c) Smart Card PHR and Mobile Apps: Smart card PHR is susceptible for being damaged or lost. In addition, in some hospitals data maybe unreadable because of unavailability of appropriate reader. Moreover, data can be inaccurate since controlled by patient. Furthermore, there is the problem of sharing data with other systems or care delivery organization when required.

d) Proposed framework: Could be susceptible for security attacks but data is encrypted and divided to many distributed parts for increased privacy. The reliance on cloud infrastructure means high ability and fault-tolerance to recover any lost data. In addition, data is shareable between all entities an allowing global access to patients' data.

Table (5) Comparison versus related work

| Name | Record Created by | Access | Patient Control | Security | Privacy | Global access | Sharing | Data entry |
|---|---|---|---|---|---|---|---|---|
| AHLTA | - | Military | low | High | Poor | Yes | Private | By doctors |
| Microsoft Health Vault | By patient | Personal access | High | High | Poor | Online access | Print paper | Patient |
| MyC² | Doctors | Doctors | No | Low | Low | Portable device | Portable device | Doctors |
| P³ HR | Patients | Patients | High | Low | Low | Portable device | Portable device | Patients |
| pPHR | Patients | Patients | High | Low | Low | Portable device | Portable device | Patients |
| MyPHR Machines | Patients | Patients | High | Low | low | Web based | Portable device | Patients |
| Dhatri | Doctors | Doctors | Low | Low | Low | Mobile based | Printed paper | Doctor |
| Mobile App | Patients | Patients | High | Low | Low | Mobile based | Portable device | Patients |
| Smart Cards | Patients | Patients | High | Low | Low | Portable device | Portable device | Patients |
| CARE (EGH)[47] | Hospital | Hospital Admins | No | Medium | Low | Local Access | Printed paper | Registered employee |
| Proposed Framework | Hospitals+ M.O.H | Doctors + Patients | High | High | High | Global user access | Shared globally | Patients + Doctors |

## 3.3 Framework Design

Adopting a cloud-based framework achieves high availability and fault-tolerance. Such issue is of paramount importance when designing a GHR access system because seconds could mean all the difference to patients.

This framework is designed to work with global health records. The actors will be divided according to the country they belong to, and each country will deploy its own private cloud for this framework which is able to communicate with other clouds under the management and control of main Cloud controlled by the W.H.O. This framework design grants each government (M.O.H.) their national (local) level of control and management. Any country not ready yet to roll out its own cloud can opt to save its pertinent data under the control of W.H.O till their cloud is up and running. Other related work discussed the national level of control [48], [49], [50], [51], and [52].

Our vision for framework infrastructure is designed adopting a multi-cloud setup per country. The first cloud of servers is considered to be *Front end cloud* for public access from all entities that connect to the framework requesting the information. Requests to the front end are relayed to another cloud of servers (*back end cloud*), as a secure encrypted request which is not accepting any communication except from the predefined Front end cloud. After the request is decrypted, the required data will be collected from the database, encrypted and sent to the Front end cloud. The front end cloud formulates the reply and resends it to the user device as shown in Figure (6). Since actors are differentiated according to their type, each type can access with different privileges and different connections, such as the predefined organization will be able to access the system through the VPN connection which will create the secure tunnel for the transferred data as shown in Figure (6).

Hospitals will be registered in the framework through M.O.H, which will create a new hospital account, and the required data for that entity such as (IPs, username, location, contact information, VPN connection Data…). When a hospital account is created, it will be able to setup a VPN connection to the framework from hospital network and manage all requests to the online domain to go through the VPN tunnel. The reason for dividing the servers to more than one cloud is to achieve more security and reduce the overhead to different framework servers. The server functions will be as described in the following subsections.

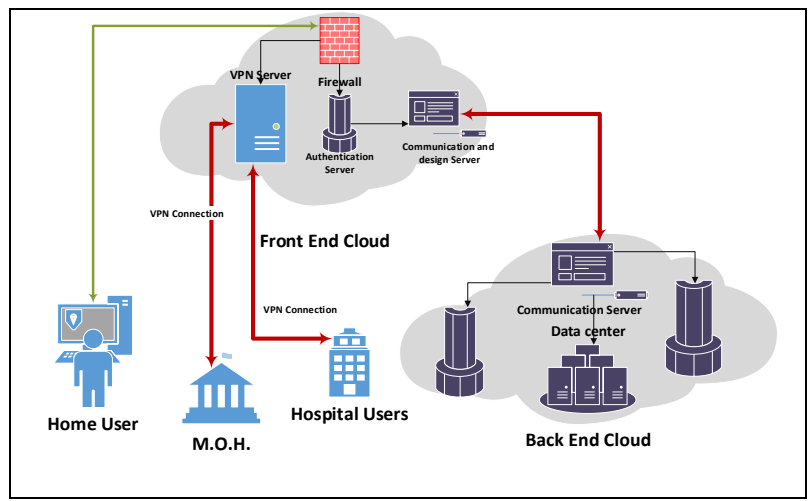

Figure (6) General view of the framework servers (clouds) per country

### 3.3.1    Front end Cloud

The main responsibility for the Front end cloud is to receive all requests from actors and generate the request to be sent to the backend cloud. The main components of the front end cloud should include a firewall, a VPN server, and related databases such as the physicians and hospitals records. In addition, it includes a response web page design generator to formulate data received from the backend cloud in a format suitable to the end user device.

The front end cloud assigns the privileges to each actor according to their type and the location they are logging in from. In addition, it performs the decryption and encryption of the transmitted data to the backend cloud. Each response from the backend cloud will be received as unformatted data which will be put in a suitable design before it can be sent to the end user device.

### 3.3.2    Backend Cloud (Database server)

The backend cloud handles patients' secure data storage. The patient data is divided into unlinked sections unless access is granted by the patient (see section 3.1.1). Only search by patient ID is allowed for physicians. This means unlinkability between records and patients unless the linkage is granted by the patient to physicians. The backend cloud responds to requests relayed by the front end cloud performing decryption and encryption as deemed necessary.

### 3.3.3 Actors Connections and Access

More than one scenario exists for actors' access to the framework for example a physician performs access from a hospital, or a physician access from anywhere (other than hospital), or organization access, or patients' access, or access to foreign patient record. In the following, each scenario will be described illustrating communication sequence.

**Organization access**: The organization needs to be registered and owns access information to establish a VPN connection. First, a log in from the registered IP for that organization is performed and then a VPN connection to the Front end cloud is attempted which will check the IP first then authentication data for validity. Upon successful authenticated connection, a VPN tunnel is created to start accessing the system data. At the organization network, any request will be routed to the VPN tunnel and the front end cloud will perform authentication.

**Physician access from connected hospital**: The hospital is already connected to the front end cloud through a VPN connection. A physician request for access will be authenticated for access by front end cloud over the secure tunnel. The front end cloud will create a token for that session including the hospital ID and physician ID. That token will be active for period of time and when it expires the physician will be requested to re-enter his authentication information. Since the token is created, the server will collect the data of that physician profile and send it to physician browser.

**Physician access from anywhere other than connected hospital**: The physician request for access will be from a dynamic IP address (un-validated IP address) and not through a VPN connection. Hence, an HTTPS connection is enforced to the front end cloud requiring proper authentication. Upon successful connection, the physician profile will be sent to the physician browser.

**Patient access**: The patient's request is directed to the front end cloud. This request is encrypted and relayed to the back end cloud. The response is formulated, decrypted, and then sent back to the front end cloud. Upon successful decryption of the response, the reply is sent back to patient profile with proper page design. When a patient visits a physician in his private clinic for diagnosis, the logged in physician will search for patient by ID then session Token, or name and session Token, or fingerprint. The Front end server will create a token for that open session to inform the backend server with the physician ID who is requesting that information, and also inform the patient with notification message that the profile was accessed by that physician with the time of access and physician location information.

**Physician access to foreign patient record**: The previous scenarios outlined the connections for local patient record access. When a logged in physician requests a foreign patient record, the front end will not recognize the ID of that patient so the requested ID will be forwarded to W.H.O. cloud which will identify the relevant cloud for that patient and redirect the request to it while notifying the requesting cloud about the location of that record. If that record is marked to be accessed globally then the requested data will be directly sent to that physician.

## 4. PROTOTYPE AND PERFORMANCE EVALUATION

A proof of concept prototype was implemented using PHP, with Apache server to clarify the main idea of the framework, for giving global access to actors, and tested with multiple physicians from different countries, getting their acceptance for the main functions. The framework provided a new way to save the patients' records, shared the PHI between patients and other health organizations, and maintained the privacy of the patient. The prototype is accessible online at http://www.Ghader.net (GHADER stands for Global Health Access for Electronic Record). For testing purposes, a local version of the prototype is deployed on a local test machine with specs Intel Core(TM)2 Duo CPU processor running at 1.83GHz, 2.00 GB RAM running Window 7 Ultimate 32-bit Operating system. Using local server (version: Apache/2.2.4 (Win32) PHP version: 5.2.3). All framework components were run locally on the test machine.

Jmeter tool [53] was used to record an actor's actions and then reload it with a number of virtual users. The actions were recorded with total 90 samples for each Jmeter user. the recorded actions were as follows: log in as hospital user, create new case for two patients describing each one's satuts then log out; and then log in again as physician 1 checking the IP address and edit one case in his department in that hospital, re-login again from another IP and create new visit for a patient in physician clinic; then log out and re-login with physician 2 who did same thing as physician 1 for another patient; then patient 1 log to system to check the cases and visits in his record and create new note explaining his status today. Finally, the second patient does the same thing as patient 1 for his/her record.

Jmeter was run with multi-users in different ramp-up seconds. The aggregate graph for testing with about 70000 samples equivalent to almost 800 users in 20 second ramp-up is shown in figure (7). The average time for requests was about 1.7 seconds. All responses for all requests were tested and all were correct and affected in the database, with 20 seconds being the maximum time spent for requests for all users.

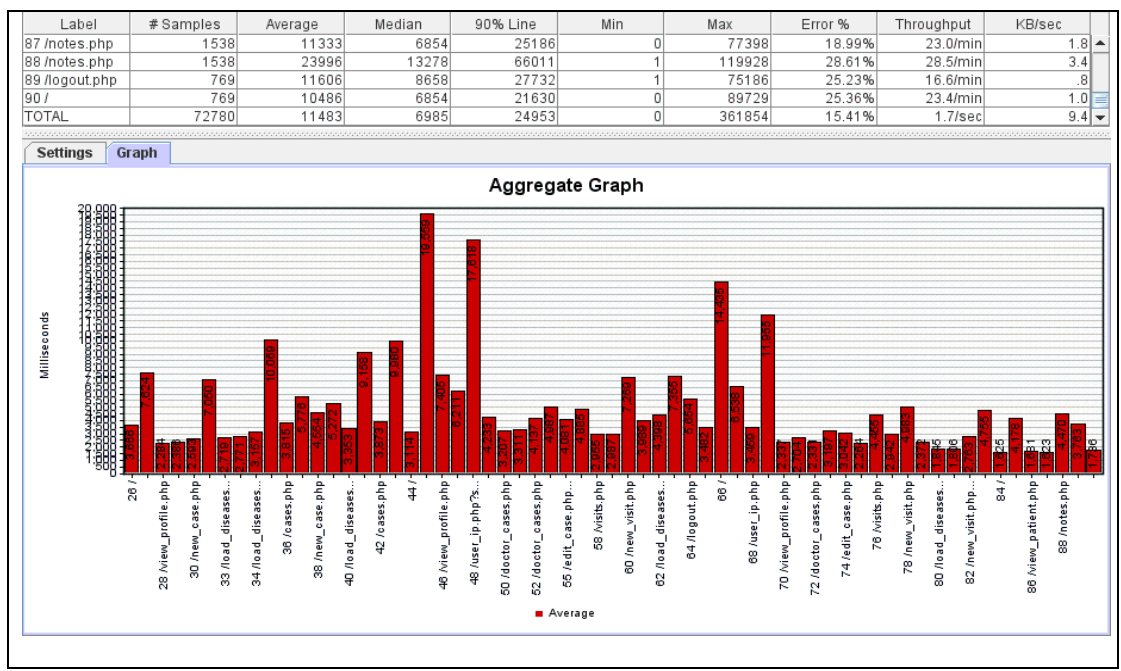

Figure (7) aggregate graph for 72780 samples run in 20 second ramp-up

Figure (8) represents the response time for different number of users accessing the implemented prototype in different ramp-up seconds on a local host. The response is different according to the number of samples. Each time the number of samples increased, the response time also increased. At the end of each test, the response time gets stable because all functions and pages were already loaded which means less load on the server.

The prototype performance evaluation results show the validity and viability of the framework. Some functions did require more time than others to get the response but in total the framework response is acceptable. In the future, a cloud-based prototype will be deployed with real users from more different countries.

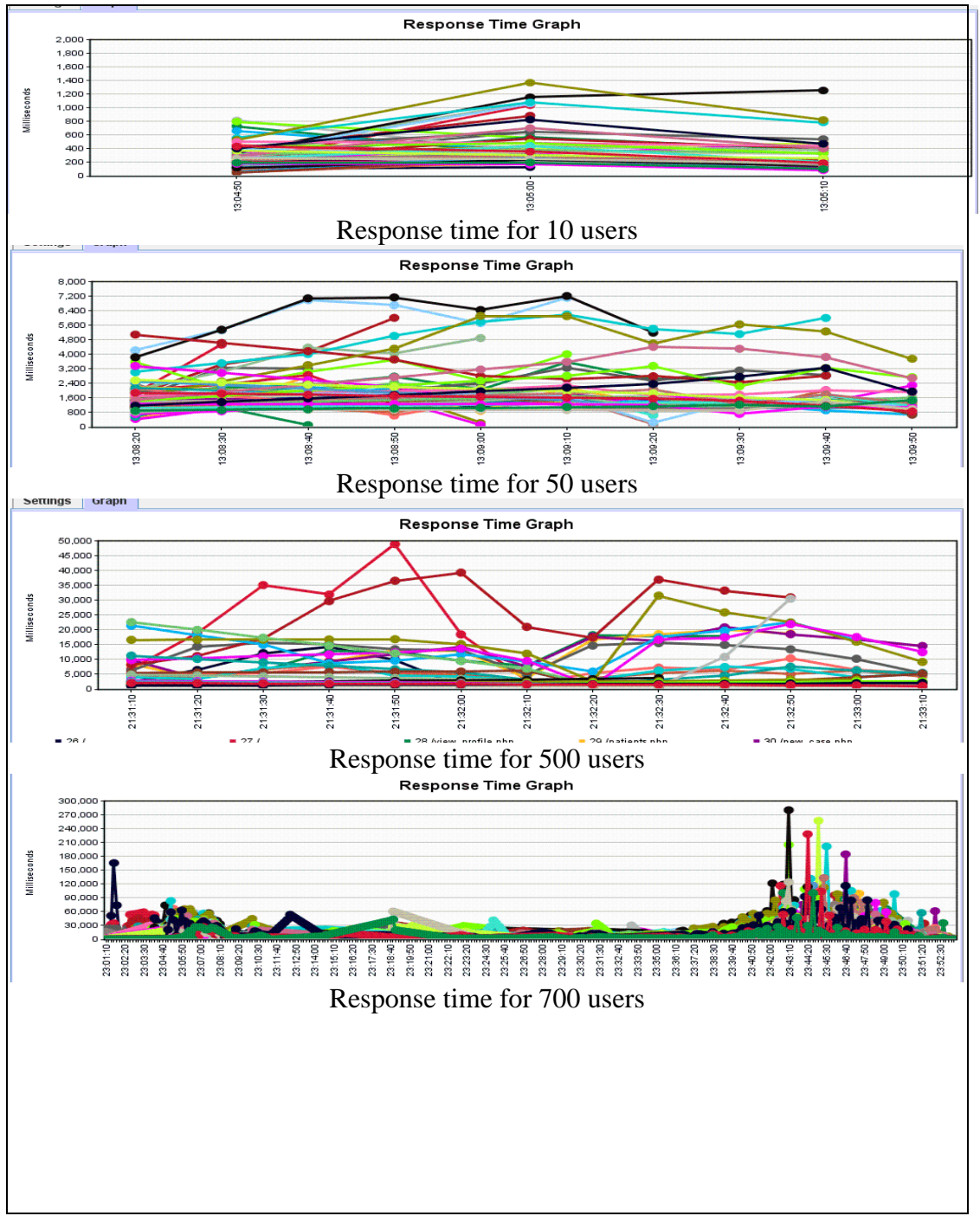

Figure (8) Multi users' response time

All previous tests were done on a local host. In some tests, the web site http://blazemeter.com/ was used for testing the online domain https://www.ghader.net and the results are as shown in Figure (9).

According to the previous tests, it is clear that the response time increases each time more users access the framework server. The previous tests included access by different type of actors each time one virtual user from JMeter is created. This created more overload on the server because some requested pages were not allowed to be open for some actors which created some errors in the response such as one patient logged in and the queue request was to view another patient's records which is allowed to be viewed only by authorized physicians, while that patient is not authorized to access this page.

Figure (10) shows the change in framework response time according to the number of samples. The response time increases when the number of samples is increased but it reaches a certain level when all services are loaded on the server and its starts decreasing due to the web pages being cached by a user's browser.

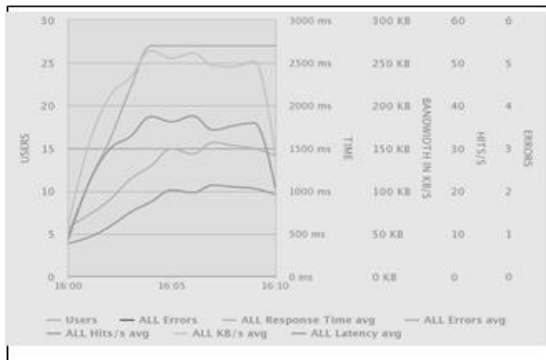

Figure (9) Blazemeter.com results with 27 users

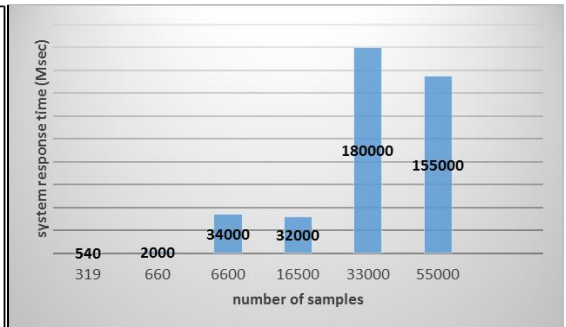

Figure (10) relation between number of samples and response time

Implementing such a framework will always face some challenges specially policy limitations. The W.H.O is recommended to be the manager for adopting the proposed framework, hence many obstacles will vanish. The only thing we envision to be remaining is the acceptance of the individual governments to implement the framework in their country and follow the framework policy and infrastructure. Another point to mention is the cost of implementation which will be high, but since each country will have their cloud infrastructure and each hospital will integrate the fingerprint or the biometric reader devices in their system, the cost will be divided between countries. Hence, we think the cost per country will be reasonable according to the benefits they will gain.

## 5. CONCLUSION AND FUTURE WORK

In this paper, we presented a novel framework for accessing the PHI of patients in a global perspective under the management of W.H.O, combining the advantages of EHR and PHR together into one record named Global Health Record "GHR". GHR grants both physicians and patients access to the patient's medical record but is mainly controlled by the patient. The proposed GHR can be easily shared and accessed by any medical entities once they are identified within the framework. Moreover, the privacy of the patient's information is granted with no misuse since the medical data is unlinkable to the patient's personal information, if not allowed by the patient himself.

According to a conducted survey [13] including 1145 physicians from 18 countries, 92% of physicians agreed to use the patient's biometric data to identify the patient and locate his/her medical record. This is because they think that it is hard to find any data about patients in some situations such as elderly patients and patients in comma [13]. Consequently, in future work we plan to implement the biometric information feature and use a realistic cloud computing setup with realistic users from many countries under the management of W.H.O.

# AUTHORS

**Nael A.H AbuOun** received his Bachelor degree in Information Technology at Sindh University, Pakistan 2004. Currently, he is working towards his master degree of information systems in the College of Computing and Information Technology, AASTMT, Alexandria, Egypt. His main research interests are Global Health Record access over cloud computing. 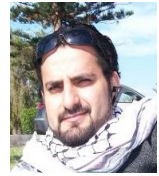

**Ayman Abdel-Hamid** is a Professor of Computer Science in the College of Computing and Information Technology, Arab Academy for Science, Technology, and Maritime Transport (AASTMT), Alexandria, Egypt. In addition, he currently holds the position of Head of Computer Science and Software Engineering Departments. He received his Ph.D. degree in Computer Science from Old Dominion University, VA, USA in May 2003. His research interests include mobile computing, network-layer mobility support, computer and network security, distributed systems, and compu ter networks. He is a member of IEEE, IEEE Computer Society, ACM, and ACM SIGCOMM. 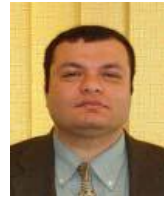

**Mohamad Abou El-Nasr** is a Professor of Computer Engineering, and Dean of Educational Affairs, Arab Academy for Science, Technology and Maritime Transport (AASTMT)—Alexandria, Egypt. He is also affiliated with Virginia Polytechnic Institute and State University where he works as an adjunct professor in the Bradley Department of Electrical and Computer Engineering - VTMENA program. He earned both his Ph.D. and M.Sc. in Electrical and Computer Engineering in March 2003 and December 1999 respectively, from Georgia Institute of Technology, Atlanta GA, USA. His research interests include UWB systems, physical MAC layer issues in wireless networks, wireless sensor networks, cloud computing, e-health, and m-health. He is a Senior Member of IEEE Communications and Computer societies, and a member of ACM. 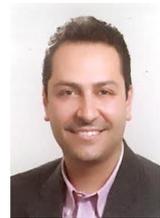